# Sub-Gbps key rate four-state continuous-variable quantum key distribution within metropolitan area

Heng Wang[1], Yang Li[1], Yaodi Pi[1], Yan Pan[1], Yun Shao[1], Li Ma[1], Jie Yang[1], Yichen Zhang[2], Wei Huang[1†] & Bingjie Xu[1]*

[1]Science and Technology on Communication Security Laboratory, Institute of Southwestern Communication, Chengdu 610041, China
[2]State Key Laboratory of Information Photonics and Optical Communications, Beijing University of Posts and Telecommunications, Beijing 100876, China

Correspondence and requests for materials should be addressed to H. W. and X. B. J. (email: †huangwei096505@aliyun.com and *xbjpku@pku.edu.cn)

Continuous-variable quantum key distribution (CVQKD) has potential advantages of high secret key rate, which is very suitable for high-speed metropolitan network application. However, the reported highest secret key rates of the CVQKD systems up to now are limited in a few Mbps. Here, we address the fundamental experimental problems and demonstrate a single-carrier four-state CVQKD with sub-Gbps key rate within metropolitan area. In the demonstrated four-state CVQKD using local local oscillator, an ultra-low level of excess noise is obtained and a high efficient post-processing setup is designed for practically extracting the final secure keys. Thus, the achieved secure key rates are 190.54 Mbps and 137.76 Mbps and 52.48 Mbps using linear channel assuming security analysis method and 233.87 Mbps, 133.6 Mbps and 21.53 Mbps using semidefinite programming security analysis method over transmission distances of 5 km, 10 km and 25 km, respectively. This record-breaking result increases the previous secret key rate record by an order of magnitude, which is sufficient to achieve the one-time pad cryptographic task. Our work shows the road for future high-rate and large-scale CVQKD deployment in secure broadband metropolitan and access networks.

Continuous-variable quantum key distribution (CVQKD) provides a secret key shared between the sender (Alice) and the receiver (Bob) with information-theoretical security [1, 2], which is very suitable for broadband metropolitan and access networks due to its inherent advantages of high key rate and good compatibility with commercial off-the-shelf components [3-5]. However, the reported CVQKD systems with several Mbps secret key rate (SKR) [6, 7] are still not up to the requirements of high-speed one-time-pad encryption (e.g. 100 Mbps secure access requirements for home networks). Therefore, the development of ultra-high SKR CVQKD is of great importance for its practical application [8, 9].

According to the modulation method of the coherent state, two practical CVKQD schemes have been proposed. One is based on Gaussian modulation coherent state (GMCS) [2, 10 and 11], and the other is based on discrete modulation coherent state (DMCS) [12-14]. The GMCS CVQKD has made great progress in both theory and experiment in recent years [15-19], (e.g., see [18] [Sec. VII and Sec. VIII] for an overview). However, high repetition rate GMCS CVQKD depends deeply on large broadband linear modulation and coherent detection, which potentially limits the SKR. As a comparison, the DMCS CVQKD, such as four-state protocol, has more practical advantages of working at low signal to noise ratio (SNR) and large operating bandwidth, which can potentially improve the SKR significantly [20-22]. Currently, high-speed DMCS CVQKD has been extensively researched by combining the local local oscillator (LLO) scheme which is free from the security loopholes and the intensity bottleneck of the transmitting LO [23-25]. However, to improve the SKR in practice, the DMCS LLO-CVQKD system faces the following issues: 1) A precise phase noise compensation (PNC) scheme is required to achieve good coherence between two independent lasers in LLO-CVQKD system [26-29]. Meanwhile, the DMCS CVQKD with large operating bandwidth needs robust approaches to eliminate other excess noises, such as photon-leakage noise, modulation and detection noise and quantization noise [28, 30]. 2) The SKRs of the reported experimental results are mostly evaluated by the linear channel assuming (LCA) security analysis method [30-33], which restricts the possible attacks performed by the eavesdropper (Eve). Therefore, a more general secure analysis theory against general collective attacks is required to estimate SKR, such as the user-defined security analysis method [11] and the semidefinite programming (SDP) method [14, 34]. 3) The demonstrated DMCS LLO-CVQKD experiments lack a high-efficient and high-speed post-processing setup to extract the final key from the raw key efficiently, which limits its practical application [35-37].

In this paper, we demonstrate a sub-Gbps key rate four-state DMCS LLO-CVQKD system experimentally within metropolitan area for the first time. In the demonstrated quantum key transceiver, the weak quantum signal and the intense pilot tone are independently generated in different optical paths, transmitted in different frequency bands and orthogonal polarization, and separately detected by two independent balanced homodyne detectors (BHDs). Compared with previous DMCS CVQKD setups, our scheme effectively reduces the modulation noise and DAC quantization noise in quantum state preparation, the photo-leakage noise in co-fiber transmission, the detection noise and ADC quantization noise in simultaneous detection. Moreover, a precise fast-slow PNC scheme to eliminate the dominate phase noise, including the pilot-tone-assisted fast-drift phase recovery and the least mean square (LMS) adaptive slow-drift phase recovery, is innovatively proposed and efficiently realized for achieving an ultra-low level of excess noise in experiment. Besides, a high efficient post-processing setup is designed to achieve rate-adaptive reconciliation efficiency better than 95% and practically extract the final secure keys in experiment. Based on the above key technological breakthroughs, the SKRs of the demonstrated CVQKD setup are 233.87 Mbps@5km, 137.76 Mbps@10km and 21.53 Mbps@25km with the SDP security analysis method and 190.54 Mbps@5km, 133.6 Mbps@10km and 52.48 Mbps@25km with the LCA security analysis method, achieving a single-carrier CVQKD with sub-Gbps key rate for the first time within metropolitan and access area.

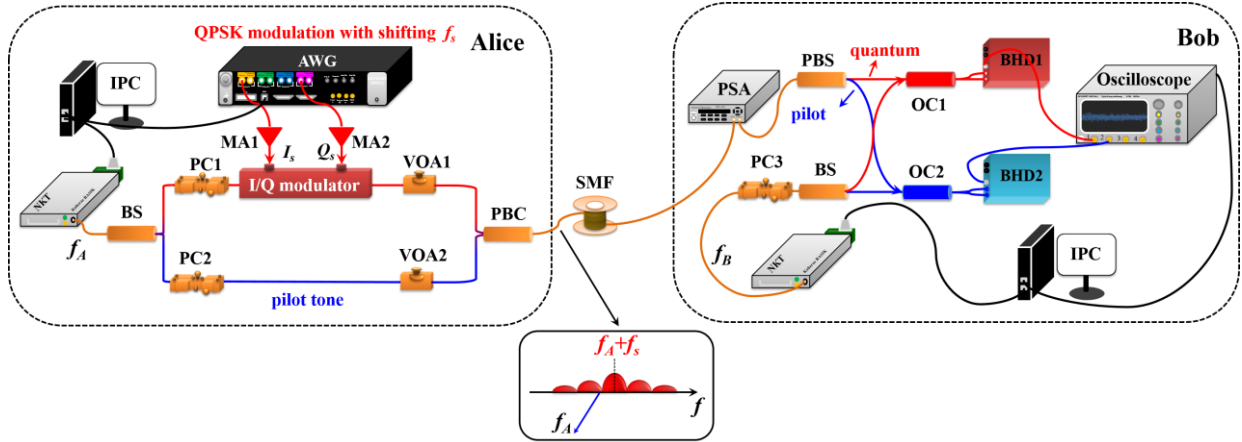

**Figure 1** Schematic setup of the proposed four-state LLO-CVQKD scheme. BS: beam splitter; PC: polarization controller; AWG: arbitrary waveform generator, MA: microwave amplifier, PSA: polarization synthesis analyzer, VOA: variable optical attenuator; PBC: polarization beam combiner; SMF: single mode fiber; PBS: polarization beam splitter; OC: optical coupler; BHD: balanced homodyne detector, IPC: industrial personal computer.

## Results

**Experimental setup.** The experimental setup of the proposed four-state discretely modulated LLO-CVQKD scheme is demonstrated in Fig. 1. At Alice's site, a continuous optical carrier is divided into two optical paths by a BS. The upper optical carrier is modulated by the QPSK digital signal with $R_{sym}$=5 GBaud symbol rate in an I/Q modulator (FUJITSU FTM7962EP), where the digital signals $I_s(t)$=real$\{[I(t)+jQ(t)]\exp(j2\pi f_s t)\}$ and $Q_s(t)$=imag$\{[I(t)+jQ(t)]\exp(j2\pi f_s t)\}$ are generated from a high-speed arbitrary waveform generator (AWG, Keysight M8195A). The QPSK modulated signal is then attenuated by a VOA to be four-state discretely modulated quantum signal. The lower optical carrier is directly attenuated to be a pilot tone with reasonable amplitude. The prepared quantum signal and pilot tone with different frequency bands and orthogonal polarization are transmitted through the quantum channel (SMF) and separated by a PBS at Bob's site. In order to separate the quantum signal and pilot tone efficiently, a PSA (General Photonics PSY-201) is used for correcting the polarization deterioration resulted from the fiber channel disturbance. Subsequently, the quantum signal and pilot tone are respectively detected with LLO signals by two commercial BHDs (Optilab BPR-23-M). In our experiment, the optical carrier at Alice's site and the LLO signal at Bob's site are independently generated from two free-running lasers (NKT Photonic Basik E15) with low RIN (<-100 dBc/Hz) and narrow linewidth (<100 Hz). So, the laser intensity noise and laser phase noise can be effectively controlled based on Eqs. (16) and (22). Moreover, two BHDs' output signals are collected and digitized by a high speed oscilloscope (Keysight DSOV084A) for the subsequent digital signal processing (DSP) and post-processing.

In the proposed four-state LLO-CVQKD system, the intense pilot tone and the weak quantum signal are independently generated in different optical paths, which is beneficial to improve the preparation accuracy of the quantum state in the case of finite modulation extinction ratio (ER) and DAC quantization bits, so that the modulation noise and DAC quantization noise can be well reduced compared with the conventional RF-subcarrier-assisted LLO-CVQKD scheme as in Eqs. (17) and (18). Moreover, the upper optical carrier is shifted by frequency $f_s$=3.5 GHz in 5 GBaud QPSK modulation relative to the lower optical carrier at Alice's site, where the photo-leakage noise from intense pilot tone to weak quantum signal can be eliminated in co-fiber transmission due to their complete isolation in frequency domain as in Eq. (19). Note that the QPSK digital signal is shaped by the root-raised cosine filter with a roll-off factor $a_{ro}$=0.3 for reducing the quantum operating bandwidth without the influence on the phase recovery accuracy in DSP. In experiment, the shifting frequency $f_s$ is mainly determined by the quantum operating bandwidth $\Delta f_q = R_{sym}*(1+a_{ro})$ and the laser frequency difference $\Delta f_{AB}$. At Bob's site, the intense pilot tone and weak quantum signal are separated in orthogonal polarization state for fully guaranteeing low-noise coherent detection of broadband quantum signal and high-saturation limitation detection of intense pilot tone. Moreover, referenced to Eq. (20) and (21), the detection noise and ADC quantization noise can be further reduced by separately detecting the intense pilot tone and weak quantum signal in the case of the limited detection dynamic and ADC quantization bits. Besides, according to Eq. (20), the low-frequency quantum noise $\varepsilon_{LF}$ can be avoided by moving the quantum signal from low-frequency band to intermediate frequency band based on heterodyne detection in our experiment. As shown in Fig. 2, the designed QPSK quantum frequency component and the designed pilot component are in different frequency bands, verifying no crosstalk between quantum signal and pilot tone. In Fig. 2a, the pilot tone is not completely suppressed due to the PBS with finite polarization isolation ratio (IR), while the residual pilot tone has no impact on the extraction of the quantum signal in completely different frequency band.

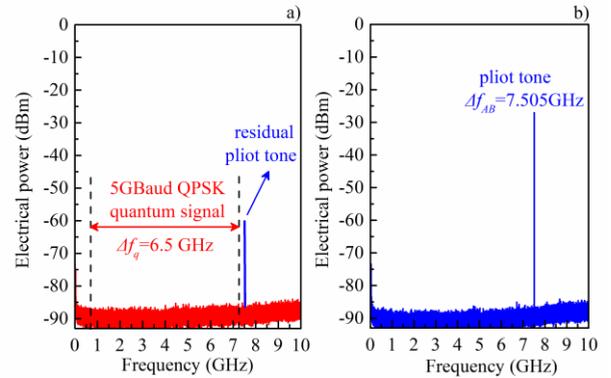

**Figure 2** Measured frequency spectra of the QPSK quantum signal and pilot tone, where $\Delta f_q$=6.5 GHz is bandwidth of the desired quantum signal and $\Delta f_{AB}$=7.505 GHz is the frequency of pilot tone. **a)** BHD1's output quantum frequency spectrum and **b)** BHD2's output pilot frequency spectrum.

**Precise fast-slow phase noise compensation.** In order to realize a reasonably low excess noise, a precise fast-slow PNC scheme is designed and realized in DSP to accurately compensate the dominate phase noise. As is illustrated in Fig. 3, the output electrical signals of two BHDs are digitized by dual-channel 8 bit ADCs at 40 GSs$^{-1}$, respectively. Firstly, the pilot tone $\Delta f_{AB}$ is precisely estimated to be 7.505 GHz by searching the peak value of the pilot frequency spectrum, and the center frequency $\Delta f_{AB}-f_s$ of the desired quantum frequency spectrum is determined to be 4.005 GHz when the shifting frequency $f_s$ is 3.5 GHz. By using the estimated frequencies, the desired quantum and pilot signals are band-pass filtered for eliminating the out-of-band noise and orthogonally down-converted for extracting the in-phase and quadrature components in baseband, respectively. Next, the baseband components of QPSK quantum signal and pilot tone are obtained by matching root-raised cosine filtering and the narrow band low-pass filtering, respectively. Note that the quantum filtering bandwidths in DSP are selected based on the detected power, the QPSK quantum symbol rate, the roll-off factor of the root-raised cosine filter and the employed laser line-width, which requires a compromise between noise suppression and phase estimation accuracy. Therefore, the fast-drift laser phase difference $\Delta \varphi_{AB}(k)$ involved in QPSK quantum signal $I_{sig}(k)+jQ_{sig}(k)$ can be recovered by sharing the phase of the pilot tone $I_{pilot}(k)+jQ_{pilot}(k)$. Moreover, the slow-drift phase difference $\Delta \varphi_d(k)$ of the QPSK quantum signal originated from different fiber delay and disturbance is adaptively recovered by the designed LMS algorithm with 51 tap and 1e$^{-3}$ step. Besides, the symbol synchronization between the transmitted and received data is finely corrected for further improving the phase recovery accuracy. Furthermore, the optical frequency difference of two free-running lasers is fixed as much as possible by precise laser wavelength control in our experiment, while the influence of small optical-frequency deviation can be eliminated by the adaptive filtering in the designed DSP. To verify the proposed DSP, the constellation diagrams of the detected QPSK quantum signal without and with the phase recovery are demonstrated respectively under transmission distance $L$=25 km, as shown in Fig. 3a and b.

**High efficient post-processing.** To extract the final key efficiently, a high efficient post-processing setup is designed as follow. Since the SNR is very low in our experiment, the raw keys after DSP, which are essentially correlated random data, are firstly reversely reconciled with the multidimensional reconciliation method [35]. After the reconciliation, the raw keys of Alice and Bob are both transferred into binary sequences, which are unidentical due to inevitable noise and will be further corrected by employing the error correction matrix based on multi-edge-type low density parity check (MET-LDPC) method

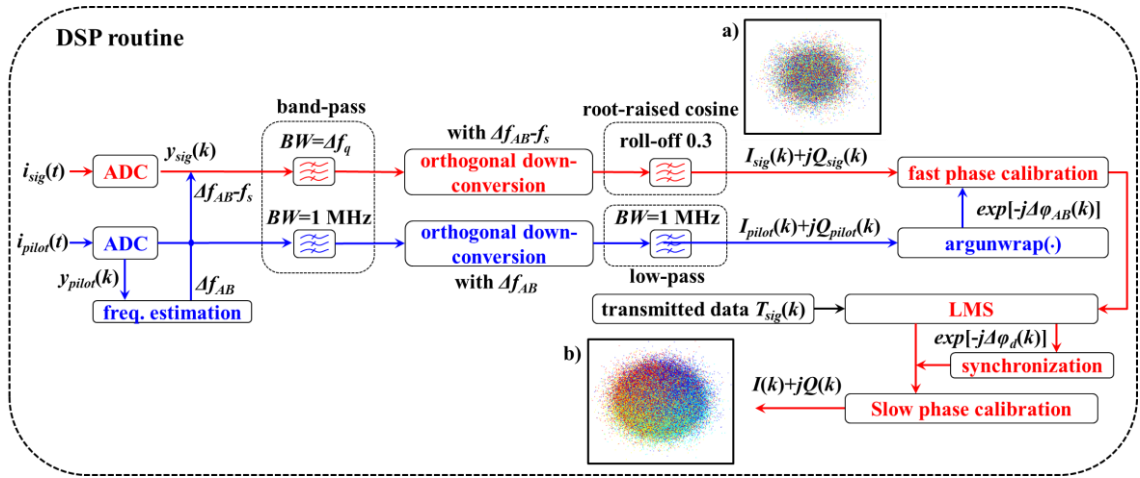

**Figure 3** The DSP routine of the proposed four-state discretely modulated LLO-CVQKD scheme, where insets **a)** without phase recovery and **b)** with phase recovery for constellation diagrams of QPSK quantum signal in the case of secure transmission distance $L$=25 km.

Table I the relative parameters of the designed high efficient post-processing

| Code rate | Degree distribution function | $\sigma^*_{DE}$ | $\beta_\sigma$ | SNR | $\beta_a$ | Distance |
|---|---|---|---|---|---|---|
| 0.07 | $v = 0.0408r_1x_1^2x_2^{28} + 0.048r_1x_1^3x_2^{29} + 0.9112r_1x_3$<br>$u = 0.0188x_1^{12} + 0.1992x_2^2x_3 + 0.712x_2^3x_3$ | 3.074 | 96.47% | 0.119 | 95.46% | 5 km |
| 0.06 | $v = 0.0522r_1x_1^2x_2^{37} + 0.0291r_1x_1^3x_2^{21} + 0.9187r_1x_3$<br>$u = 0.0213x_1^9 + 0.2136x_2^2x_3 + 0.7051x_2^3x_3$ | 3.341 | 96.94% | 0.094 | 95% | 10km |
| 0.03 | $v = 0.0249r_1x_1^2x_2^{50} + 0.0219r_1x_1^3x_2^{50} + 0.9532r_1x_3$<br>$u = 0.0105x_1^5 + 0.0063x_1^{10} + 0.5196x_2^2x_3 + 0.4336x_2^3x_3$ | 4.789 | 97.46% | 0.047 | 95.1% | 25 km |

[38-40]. Note that in order to guarantee the extraction of the final key in our experiments and hence validate the practicality of the high SKR CVQKD system proposed in this paper, the reconciliation efficiency should be achieved as high as possible. Thus, three parity check matrixes are correspondingly designed for the experiments under the transmission distance of 5 km, 10 km and 25 km with a code rate of 0.07, 0.06 and 0.03, respectively, as shown in Table I. Specially, for the design of the matrixes, a 10 bit quantization based on density evolution algorithm is chosen to obtain the degree distribution functions under such low SNRs, through which the convergence threshold $\sigma^*_{DE}$ of degree distribution function and the corresponding threshold reconciliation efficiency $\beta_\sigma$ required by the demonstrated CVQKD system is guaranteed. Subsequently, the layered LDPC decoder algorithm [36] and the adaptive decoding algorithm [37] are combined for error correction step. After the error correction, privacy amplification with Toeplitz matrix is employed to extract the final keys [41-43]. It can be observed from Table I that the threshold reconciliation efficiency $\beta_\sigma$ and the efficient rate-adaptive reconciliation efficiency $\beta_a$ are both gained to be better than 95% over the distance of 5 km, 10 km and 25 km. For our post-processing, the high efficient check matrixes are innovatively designed and efficiently realized on GPU with low SNR and final secure keys are successfully extracted in off-line situation, which are experimentally achieved in the four-state LLO-CVQKD system for the first time compared with the reported literatures according to our knowledge. Note that it is significant for high-rate CVQKD system to distribute the final secure keys between two legitimate parts by post-processing in real time, which will be deeply researched in our future work.

**Discussion**

The performance of the proposed four-state LLO-CVQKD setup is shown as follows. In our work, the SKRs of the demonstrated experimental four-state LLO-CVQKD system are firstly evaluated by the general SDP security analysis method [14] and then verified by the frequently-used LCA security analysis method [26, 32]. For achieving an optimized SKR, the SKR as a function of the excess noise and modulation variance are simulated theoretically for choosing an optimized modulation variance in applicable for the SDP and LCA method, as shown in Fig. 4. In our experiment, a preferable modulation variance $V_A$ is chosen to be about 0.45 in shot noise unit (SNU) for supporting a better SKR.

With the modulation variance $V_A$=0.456 SNU, the symbol rate $R_{sym}$=5 GBaud, the BHD's quantum efficiency $\eta$=0.45 and the BHD's electronic noise $\upsilon_{el}$=0.297, the corresponding measured excess noises are estimated on the block of size $4\times10^6$ over transmission distance $L$=5 km, 10km and 25km within continuous 30 minutes in our experiment, respectively, as in Fig. 5. The excess noise thresholds of null SKR at 5 km, 10 km and 25 km are determined to be 0.0176, 0.0141 and 0.0092 for SDP method and 0.0563, 0.0497 and 0.0371 for LCA method, respectively. Note that the tolerable excess noise by the SDP method is lower than the LCA method, which indicates that it relies on more accuracy excess noise suppression in practical setup. In our work, the laser intensity noise, modulation noise, detection noise, DAC/ADC quantization noise and photo-leakage noise from intense pilot tone to weak quantum signal can be all well eliminated by optimizing system design of the LLO-CVQKD transceiver with frequency- and polarization-multiplexing method. Moreover, a precise fast-slow PNC scheme is designed to eliminate the dominate phase noise introduced by laser phase drifting and channel phase disturbance. From Fig. 5, the mean excess noises are 0.0072, 0.0073 and 0.0075 over transmission distance of 5 km, 10 km and 25 km, respectively, achieving ultra-low level of excess noise. According to the means of measured excess noise and the obtained rate-adaptive reconciliation efficiency $\beta_a$ in Table I, the corresponding asymptotic SKRs are 233.87 Mbps/190.54 Mbps, 137.76 Mbps/133.6 Mbps and 21.53 Mbps/52.48 Mbps with the SDP/LCA methods for transmission distance of 5 km, 10 km and 25 km, respectively, as is demonstrated in Fig. 6. It is obvious that the SKRs at 5 km and 10km are increased by an order of magnitude compared with the reported single-carrier CVQKD experiments, which indicates that a sub-Gbps key rate four-state discretely modulated LLO-CVQKD is experimentally demonstrated for the first time in this paper. Notably, compared to the Ref. [14], the SKR evaluated by the LCA method is lower than that of the SDP method within transmission distance of 10 km, because the LCA method considers the trusted detection noise and electronic noise models in our work. More importantly, the ultra-low level of excess noise, the high efficient reconciliation efficiency better than 95%, and the more general secure analysis by SDP method are experimentally demonstrated in this paper, achieving high-rate and more secure four-state LLO-CVQKD system for high speed metropolitan area network application.

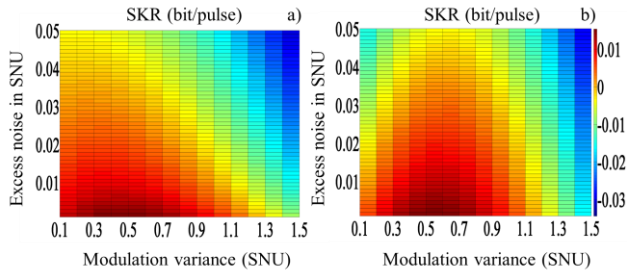

**Figure 4** Simulated thermodynamic-SKR diagram at different excess noise and modulation variance in the case of secure transmission distance of 25 km. **a)** SKR with SDP and **b)** SKR with LCA.

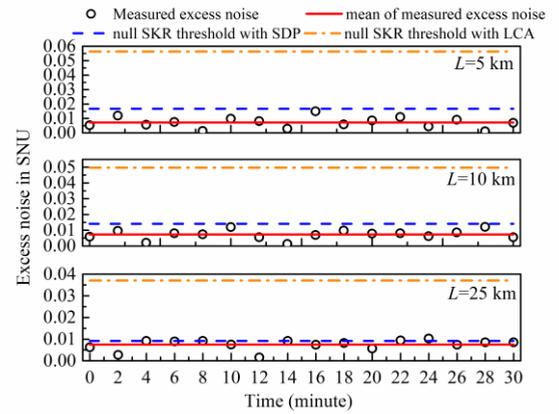

**Figure 5** Measured excess noise in SNU over 5 km, 10 km and 25 km secure transmission distance within 30 minutes. The black circles represent the measured excess noise on the block of size $4\times10^6$, the red solid lines define the mean of the measured excess noise, the blue dash lines and the orange dash dots denote the excess noises of null SKR threshold with SDP and LCA methods, respectively.

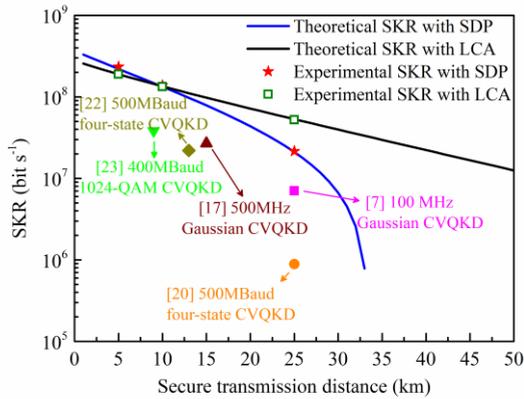

**Figure 6** SKRs as a function of the secure transmission distance. The blue and black lines represent the simulated SKRs at different secure transmission distance with the SDP and LCA method, and the red star and the olive square correspond experimental SKRs with the SDP and LCA method, respectively.

## Conclusion

We have experimentally demonstrated a sub-Gbps key rate four-state discretely modulated LLO-CVQKD scheme within metropolitan area. In the proposed scheme, the quantum signal and pilot tone are independently generated, co-propagated and separately detected based on frequency- and polarization-multiplexing method, which effectively reduces the modulation noise, ADC/DAC quantization noise, detection noise and photon-leakage noise and ADC/DAC quantization noise. Moreover, a precise fast-slow PNC scheme based on pilot-tone-assisted fast-drift phase recovery and LMS adaptive slow-drift phase recovery is designed to robustly eliminate the dominate phase noise, achieving a 5 GBaud symbol rate four-state LLO-CVQKD with an ultra-low excess noise. Furthermore, a high efficient post-processing with the rate-adaptive reconciliation efficiency better than 95% is designed to extract the final secure key experimentally (off-line), i.e. 233.87 Mbps, 137.76 Mbps and 21.53 Mbps by the SDP method and 190.54 Mbps, 133.6 Mbps and 52.48 Mbps by the LCA method over transmission distance of 5km, 10km and 25km respectively, which allows for the first time the sub-Gbps SKR single-carrier CVQKD within metropolitan area. In our work, the SDP method, which is resistant against general collective attack, is firstly used to evaluate the SKR of the experimental DMCS CVQKD setup. Moreover, the high-rate metropolitan QKD will be implemented in practice by further increasing the post-processing rate and the coherent stability of quantum key transceiver with LLO in the future. More importantly, the LLO-CVQKD with ultra-high SKR is realized to pave the way for the one-time pad encryption in secure broadband metropolitan and access networks.

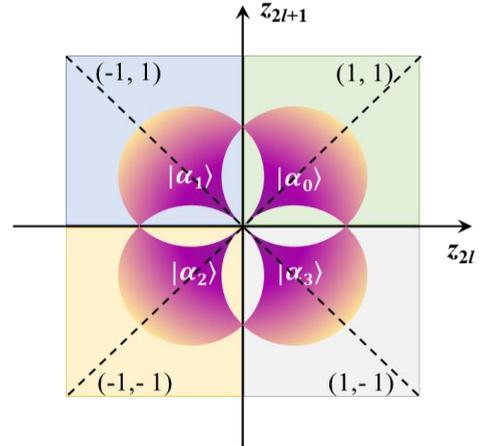

**Figure 7** Sketch map of the four coherent states protocol.

## Methods

**Four-state LLO-CVQKD protocol.** The four state CVQKD protocol can be described as follows. At Alice's site, a string of random bits $x=(x_0, …, x_{2L-1})$ are encoded as coherent states $|\psi_k\rangle$ with equal probability [14], as shown in Fig. 7

$$|\psi_k\rangle := |i^k\alpha\rangle = e^{-\alpha^2/2}\sum_{n\geq 0}e^{i(2k+1)n(\pi/4)}(\alpha^n/\sqrt{n!})|n\rangle \quad (1)$$

with $\alpha>0$ and $k\in\{0, 1, 2, 3\}$. After transmission over an insecure quantum channel, the prepared coherent states are measured by heterodyne detection at Bob's site with measurement results $z=(z_0, …, z_{2L-1})\in\mathbb{R}^{2L}$, which is converted into a raw key $y=(y_0, …, y_{2L-1})$, given by

$$(y_{2l-1}, y_{2l}) = \begin{cases} (0,0) & \text{when } z_{2l-1}\geq 0, \ z_{2l}>0 \\ (0,1) & \text{when } z_{2l-1}< 0, \ z_{2l}\geq 0 \\ (1,0) & \text{when } z_{2l-1}\leq 0, \ z_{2l}< 0 \\ (1,1) & \text{when } z_{2l-1}> 0, \ z_{2l}\leq 0 \end{cases} \quad (2)$$

Then, the parameter estimation is performed to calculate how much secret key can be achieved from the raw key via post-processing. The post-processing process includes reverse reconciliation, error correction and privacy amplification. In the

asymptotic limit, the SKR with reverse reconciliation can be written as

$$R = R_{sym}(\beta I_{AB} - S_{BE}) \quad (3)$$

where $I_{AB}$ is the Shannon mutual information between Alice and Bob, and $S_{BE}$ is the Holevo bound between Bob and Eve, respectively. Currently, the security proofs of CVQKD with four-state modulation have been established by LCA method [32, 33] and SDP method [14, 34]. In most reported four-state LLO-CVQKD experiments, the LCA method is used to evaluate the SKR [26], which limits the attack of Eve. Meanwhile, the SDP method is applicable for general collective attacks. However, the tolerable excess noise with SDP method is very low (e.g. 0.01 in shot noise unit), which is challenging in practical CVQKD system. In our work, the SDP method is verified experimentally for the first time with reasonably low excess noise.

**Security analysis with LCA method.** For the LCA method, the $I_{AB}$ in Eq. (3) can be expressed as [26, 32]

$$I_{AB} = \log_2 \frac{V + \chi_{line} + \chi_{het}/T}{1 + \chi_{line} + \chi_{het}/T} \quad (4)$$

with

$$V = V_A + 1 \quad (5a)$$
$$\chi_{line} = 1/T - 1 + \varepsilon \quad (5b)$$
$$\chi_{het} = [(2-\eta) + 2\upsilon_{el}]/\eta \quad (5c)$$

with the transmittance efficiency $T$, the modulation variance $V_A = 2\alpha^2$ and the excess noise $\varepsilon$. At the same time, $S_{BE}$ can be calculated as

$$S_{BE} = G\left(\frac{\lambda_1 - 1}{2}\right) + G\left(\frac{\lambda_2 - 1}{2}\right) - G\left(\frac{\lambda_3 - 1}{2}\right) - G\left(\frac{\lambda_4 - 1}{2}\right) \quad (6)$$

where the Von Neuman entropy $G(x) = (x+1)\log_2(x+1) - x\log_2(x)$, and symplectic eigenvalues $\lambda_i$ can be derived from the covariance matrix between Alice and Bob, which are expressed as,

$$\lambda_{1,2} = \sqrt{\frac{1}{2}\left(A \pm \sqrt{A^2 - 4B}\right)} \quad (7a)$$

$$\lambda_{3,4} = \sqrt{\frac{1}{2}\left(C \pm \sqrt{C^2 - 4D}\right)} \quad (7b)$$

with

$$A = V^2 + T^2(V + \chi_{line})^2 - 2TZ_4^2 \quad (8a)$$

$$B = (TV^2 + TV\chi_{line} - TZ_4^2)^2 \quad (8b)$$

$$C = \frac{A\chi_{het}^2 + B + 1 + 2\chi_{het}[V\sqrt{B} + T(V + \chi_{line})] + 2TZ_4^2}{[T(V + \chi_{line} + \chi_{het}/T)]^2} \quad (8c)$$

$$D = \frac{(V + \chi_{het}\sqrt{B})^2}{[T(V + \chi_{line} + \chi_{het}/T)]^2} \quad (8d)$$

$$Z_4 = 2\alpha^2 \left(\xi_0^{3/2}\xi_1^{-1/2} + \xi_1^{3/2}\xi_2^{-1/2} + \xi_2^{3/2}\xi_3^{-1/2} + \xi_3^{3/2}\xi_0^{-1/2}\right) \quad (8e)$$

where $\xi_{0,2} = 1/2\exp(-\alpha^2)[\cosh(\alpha^2)\pm\cos(\alpha^2)]$ and $\xi_{1,3} = 1/2\exp(-\alpha^2)[\sinh(\alpha^2)\pm\sin(\alpha^2)]$.

**Security analysis with SDP method.** For the SDP method, the $I_{AB}$ in Eq. (3) is expressed as [14]

$$I_{AB} = \log_2\left(1 + \frac{2T\alpha^2}{2 + T\varepsilon}\right) \quad (9)$$

where we have defined the quantum efficiency $\eta = 1$ and the electronic noise $\upsilon_{el} = 0$. The Holevo bound $S_{BE}$ can be simplified as

$$S_{BE} = G\left(\frac{v_1 - 1}{2}\right) + G\left(\frac{v_2 - 1}{2}\right) - G\left(\frac{v_3 - 1}{2}\right) \quad (10)$$

where $v_3 = 1 + 2\alpha^2 - [Z^{*2}/(1+\upsilon)]$. $v_1$ and $v_2$ are the symplectic eigenvalues of the optimized covariance matrix between Alice and Bob, given by

$$\Gamma^* = \begin{bmatrix} (1+2\alpha^2)II_2 & Z^*\sigma_z \\ Z^*\sigma_z & \upsilon II_2 \end{bmatrix} \quad (11)$$

where $\upsilon = 1 + 2T\alpha^2 + T\varepsilon$. $II_2 = diag[1, 1]$ and $\sigma_z = diag[1, -1]$ are the diagonal matrices. $Z$ is the optimal solution of the following constraint condition

$$\min \operatorname{tr}\left[\left(\prod a \prod \otimes b + \prod a^\dagger \prod \otimes b^\dagger\right) X\right]$$

$$\begin{cases} \operatorname{tr}\left\{\left[\prod \otimes (1+2b^\dagger b)\right] X\right\} = \upsilon \\ \operatorname{tr}\left\{\begin{bmatrix} (|\psi_0\rangle\langle\psi_0| - |\psi_2\rangle\langle\psi_2|) \otimes \hat{q} \\ +(|\psi_1\rangle\langle\psi_1| - |\psi_3\rangle\langle\psi_3|) \otimes \hat{p} \end{bmatrix} X\right\} = 2\sqrt{T}\alpha \\ \operatorname{tr}(B_{k,\ell} X) = \frac{1}{4}\langle\alpha_\ell | \alpha_k\rangle \\ X \succeq 0 \end{cases} \quad (12)$$

with the annihilation and creation operators $a$ ($b$) and $a^\dagger$ ($b^\dagger$) on Fock space at Alice's site and Bob's site, respectively. $X$ is positive semidefinite. We have defined $B_{\ell,k} = |\psi_\ell\rangle\langle\psi_k|$ ($\ell, k = 0, 1, 2, 3$) and $\Pi = |\psi_0\rangle\langle\psi_0| + |\psi_1\rangle\langle\psi_1| + |\psi_2\rangle\langle\psi_2| + |\psi_3\rangle\langle\psi_3|$. In the SDP method, the four coherent states can be expressed as

$$|\psi_k\rangle = \frac{1}{2}\sum_{m=0}^{3} e^{-j(2k+1)m\pi/4}|\phi_m\rangle \quad (13)$$

where

$$|\phi_m\rangle = \frac{1}{\sqrt{\xi_m}}\sum_{n=0}^{\infty}\frac{\alpha^{4n+m}}{\sqrt{(4n+m)!}}|4n+m\rangle \quad (14)$$

with $\xi_{0,2} = 1/2[\cosh(\alpha^2)\pm\cos(\alpha^2)]$ and $\xi_{1,3} = 1/2[\sinh(\alpha^2)\pm\sin(\alpha^2)]$.

**Excess noise model of four-state LLO-CVQKD setup.** For distilling the final key with the above-mentioned LCA and SDP methods, the designed four-state LLO-CVQKD system relies on a low level of excess noise. In the four-state LLO-CVQKD scenario, several main excess noise components are considered and modeled as [30]

$$\varepsilon = \varepsilon_{RIN} + \varepsilon_{DAC} + \varepsilon_{Mod} + \varepsilon_{LE} + \varepsilon_{Det} + \varepsilon_{ADC} + \varepsilon_{Phase} \quad (15)$$

The first term $\varepsilon_{RIN}$ represents the laser intensity noise of two independent lasers, which mainly includes two parts

$$\varepsilon_{RIN} = TV_A\sqrt{RIN_{quan}B} + \frac{1}{4}RIN_{LO}BV_{RIN}(\hat{q}) \quad (16)$$

where $RIN_{quan}$ and $RIN_{LO}$ are the relative intensity noises of the signal laser and the LO, respectively. $B$ denotes the electronic bandwidth and $V_{RIN}(\hat{q})$ describes the quantum variance without taking the LO's relative intensity noise (RIN) into account.

The second term $\varepsilon_{DAC}$ in Eq. (15) is the quantization noise introduced by the additional voltage error of the quadratures of the signal in finite DAC quantization bits, given by

$$\varepsilon_{DAC} \leq TV_A \left[ \pi \frac{\delta V_{DAC}}{V_{DAC}} + \frac{\pi^2}{2} \left( \frac{\delta V_{DAC}}{V_{DAC}} \right)^2 \right]^2 \quad (17)$$

where $V_{DAC}$ is voltage translated from the signal-bit information, and the deviation voltage $\delta V_{DAC}$ is determined by the quantization bits and full voltage range of DCA.

In the conventional pilot assisted LLO-CVQKD scenario, the modulation noise $\varepsilon_{Mod}$ in Eq. (15) can be expressed as [44]

$$\varepsilon_{Mod} = |a_R|^2 10^{-d_{dB}/10} \quad (18)$$

where $d_{dB}$ represents the ER of the employed I/Q modulator and $a_R$ means the amplitude of the intense pilot pulse. It is obvious from Eq. (17) and (18) that the quantization noise $\varepsilon_{DAC}$ and modulation noise $\varepsilon_{Mod}$ would be reduced in finite DAC quantization bits and modulation ER when the weak quantum signal and intense pilot tone are separately generated in different modulation path in contrast with the reported RF-subcarrier-assisted LLO-CVQKD scheme [26, 28].

The fourth term $\varepsilon_{LE}$ in Eq. (15) denotes the photon-leakage noise, which is determined as [44, 45]

$$\varepsilon_{LE} = \frac{2|a_R|^2}{R_e} \quad (19)$$

where $R_e$ denotes the IR, and it mainly depends on polarization IR and modulation ER when the quantum signal and pilot tone are in same time duration or frequency band based on time multiplexing or frequency multiplexing. In the former LLO-CVQKD schemes, the surplus pilot signals cannot be completely suppressed due to finite modulation ER and polarization IR, resulting in photon-leakage noise on the quantum signal in practical experiment [7, 44]. Therefore, it is better to completely isolate the quantum signal and pilot tone in frequency or time domain for eliminating the photon-leakage noise.

From Eq. (15), the fifth term $\varepsilon_{Det}$ represents the heterodyne detection noise at Bob's site, given by

$$\varepsilon_{Det} = 2 \frac{NEP^2}{hf} \frac{B\tau}{P_{LO}} + \varepsilon_{LF} \quad (20)$$

with the noise-equivalent power $NEP$ and Planck's constant $h$. $f$ and $P_{LO}$ are the LO frequency and power, respectively. $\tau$ denotes the pulse duration. Note that an additive noise $\varepsilon_{LF}$ denotes the low-frequency quantum noise, which is mainly determined by the low-frequency linearity of BHD and the linewidth of laser.

In the heterodyne detection case, the ADC quantization noise $\varepsilon_{ADC}$ in Eq. (15) can be expressed as

$$\varepsilon_{ADC} = 2 \frac{\tau}{hfg^2 \rho^2 P_{LO} \eta T} \frac{1}{12} \frac{R_U^2}{2^{2n}} \quad (21)$$

where $g$ is the gain factor of the amplifier (in V/A) and $\rho$ is the responsivity of the PIN diodes (in A/W). $R_U$ and $n$ are the full voltage range and quantization bits of ADC, respectively. It is obvious that the ADC quantization noise will be alleviated in finite quantization bits if the weak quantum signal and intense pilot tone are independently quantized by two ADCs at Bob's site for separate detection. Moreover, we can see from Eq. (20) and Eq. (21) that the separate detection can flexibly provide the sufficient optical power $P_{LO}$ to reduce the detection noise and quantization noise as much as possible in finite detection dynamic and ADC quantization bits.

Finally, from Eq. (15) that the last term $\varepsilon_{Phase}$ represents the dominate phase noise in excess noise. In LLO-CVQKD scenario, the phase noise is divided into two parts, given by

$$\varepsilon_{Phase} = \varepsilon_{phase,fast} + \varepsilon_{phase,slow} \quad (22)$$

where the fast-drift phase noise $\varepsilon_{phase,fast}$ is originated from the relative phase drift between two independent lasers, determined by $\varepsilon_{phase,fast} = 2\pi V_A (\Delta v_A + \Delta v_B)/R_{sym}$. So, two independent laser linewidths $\Delta v_A$ and $\Delta v_B$ should be chosen to be small as much as possible in experiment. Moreover, the slow-drift phase noise $\varepsilon_{phase,slow}$ maybe come from the fiber channel disturbance and the phase difference between quantum signal and pilot tone in the pilot assisted channel equalization recently applied in optical fiber LLO-CVQKD [46, 47]. In our work, a precise fast-slow PNC scheme is proposed based on pilot-tone-assisted fast-drift phase recovery and LMS adaptive slow-drift phase recovery for achieving an ultra-low level of excess noise.

**Data availability**
The Data used in this study are available from the authors under reasonable request.

**Code availability**
Code used in the study is available from the authors under reasonable request.

**Acknowledgments**
We acknowledge the financial support from the National Science Foundation of China (Grants No. 62101516, No. 62171418, No. 61771439, No. U19A2076, and No. 61901425), the Chengdu Major Science and Technology Innovation Program (2021-YF08- 00040-GX), the Technology Innovation and Development Foundation of China Cyber Security (Grants No. JSCX2021JC001), the Sichuan Application and Basic Research Funds (Grants No. 2020YJ0482 and No. 2021YJ0313), the Sichuan Science and Technology Program (Grants No. 2019JDJ0060 and No. 2020YFG0289), the National Cryptography Development Fund (Grant No. MMJJ20170120).

**Author contributions**
W. H. proposed the idea and wrote this manuscript. W. H, P. Y. D. and P. Y. carried out the experimental work. S. Y. and H. W. carried out the excess noise modeling. M. L., Y. J. and L. Y. carried out the post-processing work. Z. Y. C. and X. B. J. carried out the theoretical analysis on the protocol. All the authors analyzed and discussed the results and contributed to write the manuscript.

**Competing interests**
The authors declare no competing interests.